\documentclass[fleqn,usenatbib]{mnras}

\usepackage{newtxtext,newtxmath}

\usepackage[T1]{fontenc}
\usepackage{ae,aecompl}

\usepackage{graphicx}    
\usepackage{amsmath}    
\usepackage{amssymb}    
\usepackage{amssymb}    
\usepackage{amsbsy}    

\usepackage[usenames]{color}
\definecolor{verdone}{rgb}{0,0.398,0}

\title[The surface effect on the RGB]{Investigating surface correction relations for RGB stars}

\author[A. C. S. J{\o}rgensen et al.]{%
Andreas Christ S{\o}lvsten J{\o}rgensen$^{1}$\thanks{E-mail: a.c.s.joergensen@bham.ac.uk},
Josefina Montalb{\'a}n$^{1}$,
Andrea Miglio$^{1}$,
\and
Ben M. Rendle$^{1}$,
Guy R. Davies$^{1}$,
Ga{\"e}l Buldgen$^{2}$,
Richard Scuflaire$^{3}$,
Arlette Noels$^{3}$,
\and
Patrick Gaulme$^{4}$ 
and Rafael A. Garc{\'i}a$^{5,6}$
\\
$^{1}$School of Physics and Astronomy, University of Birmingham, Edgbaston B15 2TT, UK \\
$^{2}$Observatoire de Geneve, Universit{\'e} de Gen{\`e}ve, 51 Ch. Des Maillettes, CH-1290 Sauverny, Suisse \\
$^{3}$Space sciences, Technologies and Astrophysics Research (STAR) Institute,
Universit{\'e} de Li{\`e}ge, 19C All {\'e}e du 6 A{\^o}ut,B-4000 Li{\`e}ge, Belgium \\
$^{4}$Max-Planck-Institut f{\"u}r Sonnensystemforschung, Justus-von-Liebig-Weg 3, 37077 G{\"o}ttingen, Germany \\
$^{5}$IRFU, CEA, Universit{\'e} Paris-Saclay, F-91191 Gif-sur-Yvette, France \\
$^{5}$AIM, CEA, CNRS, Universit{\'e} Paris-Saclay, Universit{\'e} Paris Diderot, Sorbonne Paris Cit{\'e}, F91191 Gif-sur-Yvette, France
\\
}

\date{Accepted XXX. Received YYY; in original form 24.04.2020}

\pubyear{2020}

\begin{document}
\label{firstpage}
\pagerange{\pageref{firstpage}--\pageref{lastpage}}
\maketitle


\begin{abstract}
State-of-the-art stellar structure and evolution codes fail to adequately describe turbulent convection. For stars with convective envelopes, such as red giants, this leads to an incomplete depiction of the surface layers. As a result, the predicted stellar oscillation frequencies are haunted by systematic errors, the so-called surface effect. Different empirically and theoretically motivated correction relations have been proposed to deal with this issue. In this paper, we compare the performance of these surface correction relations for red giant branch stars. For this purpose, we apply the different surface correction relations in asteroseismic analyses of eclipsing binaries and open clusters. In accordance with previous studies of main-sequence stars, we find that the use of different surface correction relations biases the derived global stellar properties, including stellar age, mass, and distance estimates. We furthermore demonstrate that the different relations lead to the same systematic errors for two different open clusters. Our results overall discourage from the use of surface correction relations that rely on reference stars to calibrate free parameters. Due to the demonstrated systematic biasing of the results, the use of appropriate surface correction relations is imperative to any asteroseismic analysis of red giants. Accurate mass, age, and distance estimates for red giants are fundamental when addressing questions that deal with the chemo-dynamical evolution of the Milky Way galaxy. In this way, our results also have implications for fields, such as galactic archaeology, that draw on findings from stellar physics.
\end{abstract}




\section{Introduction}

Asteroseismology, i.e. the study of stellar oscillations, has become an invaluable tool for stellar physics, yielding unique insights into stellar structures as well as precise stellar parameters \citep[e.g.][]{2010aste.book.....A,Chaplin2013}. This makes asteroseismology the backbone of other flourishing disciplines, branching from exoplanet research to galactic archaeology \citep[e.g.][]{Miglio2009,2010ApJ...713L.164C,2011ApJ...729...27B,2012ASSP...26...11M,2013Sci...342..331H,Miglio2013,2013ApJ...774L..19V,Lundkvist2018,2019LRSP...16....4G}.

The success of asteroseismology relies on comparisons between state-of-the-art stellar models with observations. However, to give a holistic depiction of stellar structures and their evolution, current stellar models draw on a set of simplifying approximations that result in prominent tensions with data, which greatly complicates such comparisons. One prevailing approximation is the use of mixing length theory \citep[MLT][]{Bohm-Vitense1958} or similar parameterizations to model turbulent convection. Another such simplification is the assumption that stars are spherically symmetric. In tandem, these approximations result in an incorrect depiction of the outermost, superadiabatic layers of stars with convective envelopes. This model inadequacy affects predicted model frequencies that, therefore, show a systematic offset relative to observations. The described frequency shift is known as the structural surface effect. 

By drawing upon multi-dimensional, radiative magneto-hydrodynamic (MHD) simulations of convection \citep[e.g.][]{Trampedach2013,Magic2013}, it has been shown that the structural deficiencies that underlie the surface effect can be successfully overcome. This has been demonstrated by different authors and has been accomplished during the post-processing of stellar structure models \citep{Rosenthal1999,Piau2014,Sonoi2015, Ball2016, Magic2016, Joergensen2017,Manchon2018,Joergensen2019}. 
Furthermore, recently, stellar models that use information from multi-dimensional MHD simulations throughout the stellar evolution and hereby successfully mimic the structure of such simulations have become available \citep{Joergensen2018,JoergensenWeiss2019,JoergensenAngelou2019,Mosumgaard2020}.

In addition to the structural surface effect, the frequencies of stellar models suffer from the so-called modal surface effect, denoting yet another systematic frequency offset. The modal surface effect stems from the fact that oscillation frequencies of stars are computed under the assumption of adiabaticity, which does not hold true in the near-surface layers \citep[e.g.][]{2004ASPC..310..458D,Christensen-Dalsgaard2008a,2013MNRAS.435.3406T,2014A&A...572A..11G}. Moreover, the contribution from turbulent pressure, i.e. the pressure stemming from the bulk motion of a convective fluid, is often not correctly accounted for, further contributing to the aforementioned frequency shift \citep[cf.][for a detailed discussion of this issue]{Houdek2017,Houdek2019,JoergensenWeiss2019,Schou2020}.

To compare model frequencies with observations, many authors rely on empirical surface correction relations that address the combined surface effect. The first such relation was proposed by \cite{Kjeldsen2008}. The relation amounts to a power law, for which \cite{Kjeldsen2008} calibrated the involved free parameters to mitigate the surface effect of the present-day Sun. However, there is no physical justification for using a power law, and the necessity of calibrating the relation makes it unsuitable for any star whose global parameters deviate strongly from those of the reference star \citep[cf.][]{Ball2017,Joergensen2019}. To mitigate these drawbacks, one may include a large sample of reference stars to evaluate how the power-law description evolves across the Hertzsprung-Russel (HR) diagram and investigate different functional forms. This was accomplished by \cite{Sonoi2015}, who employed three-dimensional MHD simulations to overcome the structural inadequacies of stellar models. 

Like the surface correction relation by \cite{Kjeldsen2008}, the surface correction relation by \cite{Sonoi2015} is, of course, still subject to a selection bias: both surface correction relations can only be applied to stars, whose global parameters are similar to those of the employed reference stars or models \citep{Joergensen2019}. This being said, \cite{Sonoi2015} cover the global parameters of many main-sequence stars in the \textit{Kepler} field \citep{Borucki2010x}. The surface correction relation by \cite{Sonoi2015} meanwhile suffers from a different drawback: the computation of the reference model frequencies relies on the so-called gas $\Gamma_1$ approximation \citep{Rosenthal1999}. This means that the contributions of the turbulent pressure and non-adiabatic effects to the reference frequencies are not correctly accounted for. While the gas $\Gamma_1$ approximation recovers the observed frequencies reasonably well\footnote{Albeit, even in the case of the Sun, the gas $\Gamma_1$ approximation only recovers the observed frequencies within a few microhertz.}, in the case of the present-day Sun, it is unclear how this approximation performs for any other star.

In contrast to \cite{Kjeldsen2008} and \cite{Sonoi2015}, \cite{Ball2014} present a surface correction relation with a physically motivated functional form based on an asymptotic analysis by \cite{1990LNP...367..283G}. While their surface correction relation still includes free parameters, these can be adjusted anew for each target star, circumventing the need for calibrating the parameters based on reference stars. 

As regards main-sequence stars, studies by \cite{Basu2018} and \cite{Nsamba2018} show that the use of the surface correction relation by \cite{Ball2014} leads to estimates for the stellar mass, radius, and age that are consistent with those obtained from alternative methods for handling the surface effect. One such alternative approach relies on a set of frequency ratios that were originally proposed by \cite{Roxburgh2003} and that have been shown to be insensitive to the incorrect depiction of the near-surface layers \cite{Oti2005}. Studies by \cite{Ball2017} and \cite{Nsamba2018} furthermore show that the use of the surface correction relations by \cite{Kjeldsen2008} and \cite{Sonoi2015} introduce systematic errors in the stellar parameter estimates when addressing main-sequence stars and subgiants. How well the surface correction relation by \cite{Ball2014}, or indeed any of the relations and methods mentioned above, perform throughout the HR diagram, including more evolved stages, is yet to be settled \citep[see also][]{Ball2018}. 
In this paper, we address this issue by investigating how the use of different surface correction relations affect stellar parameter estimates for red giants. 

For this purpose, we derive stellar parameter estimates adopting different surface correction relations for eight well-constrained eclipsing binaries as well as 19 and 30 red giants in the open clusters NGC~6819 and NGC~6791, respectively \citep[e.g.][]{2008A&A...492..171G,Basu2011,Brogaard2012,Miglio2012,2013AJ....146...58J,Sandquist2013,Brogaard2015,Bossini2017,Handberg2017,Rodrigues2017,2019ApJ...874..180M}.

Accurate parameter estimates for the ages of red giants are essential for establishing the dynamical and chemical evolution of the Milky Way galaxy. The performance of asteroseismology for evolutionary stages beyond the main sequence thus has profound implications for galactic archaeology, making the present study a valuable stepping stone for future analyses. {Red giants are furthermore of interest to exoplanet research when addressing the dynamics and fate of planetary systems.}

Section~\ref{sec:AIMS} addresses the underlying MCMC approach as well as the employed stellar models. We introduce the different surface correction relations in Section~\ref{sec:surfcorr}. In Section~\ref{sec:EB}-\ref{sec:discussion7}, we successively present the analyses of the eclipsing binaries and the two clusters. Our main conclusions are summarized in Section~\ref{sec:conclusion}. 

\section{\textsc{AIMS}: Bayesian inference of stellar parameters} \label{sec:AIMS}

To evaluate posterior distributions of stellar parameters, we compare model predictions with observations for each individual star. For this purpose, we employ the open-source code, Asteroseismic Inference on a Massive scale \citep[\textsc{aims},][]{Reese2016,LundReese2018,2019MNRAS.484..771R}, which is based on the Markov Chain Monte Carlo (MCMC) ensemble sampler by \cite{Goodman2010} using the implementation by \cite{emcee}.

A large variety of Monte Carlo methods have found their way into modern astrophysics. They have yielded new insights and robust parameter estimates in {\color{black}a variety of astrophysical fields and analyses}, ranging from the peak bagging of stellar oscillation frequencies to cosmology \citep[e.g.][]{2009A&A...506.1043G,Handberg2011,Porqueres2019,Porqueres2019Lya}. Indeed, the employed oscillation frequencies have been derived from the observed light curves using an open-source MCMC peak-bagging algorithm called PBjam\footnote{Cf. \url{https://github.com/grd349/PBjam}} that likewise builds on the MCMC ensemble sampler by \cite{Goodman2010}. The observed frequencies were subsequently corrected to account for the Doppler shift that arises from the line-of-sight motion of the star relative to the observer \citep{2014MNRAS.445L..94D}. {Three of the eight eclipsing binaries constitute exceptions: the mode identification of KIC~4054905, KIC~4663623, and KIC~9540225 was performed using a maximum a posteriori method based on \cite{Gaulme2009} and Benbakoura et al. (submitted). }

However, due to the high computational cost of successively computing a series of stellar models, Monte Carlo algorithms are not widespread in asteroseismic analyses that seek to derive stellar parameters. Some notable exceptions meanwhile exist, including analyses of the present-day Sun \citep{Bahcall2006,Joergensenjcd2017,Vinyoles2017} as well as a handful of other main-sequence benchmark stars \citep[e.g.][]{Benomar2009,Bazot2012,JoergensenAngelou2019}.
To bypass the high computational cost of MCMC, \textsc{aims} computes new samples for the Markov chain, i.e. models with a new combination of global stellar parameters, by interpolation in an existing grid of stellar models (cf. Section~\ref{sec:grid}). In this way, \textsc{aims} is able to compute a large set of samples with a total computation time of a few hours: for each star, we compute 2000 samples for each of the 800 walkers with 10 different temperatures \citep[see][for a general introduction to MCMC algorithms]{Gregory}. These samples are preceded by a burn-in phase of 4000 samples per walker. Based on thousands or even millions of samples, \textsc{aims} is thus able to robustly map the posterior probability distributions for the stellar parameters of evolved stars, in the same time as it takes to run a handful of stellar evolution models from the zero-age main sequence (ZAMS) to the red giant branch (RGB).

\subsection{Stellar model grid} \label{sec:grid}

We constructed two grids of stellar models on the RGB, using the \textsc{cl{\'e}s} \citep[Code Li{\'e}geois d'{\'Evolution Stellaire;}][]{CLES} stellar evolution code, and computed the associated adiabatic model frequencies, using LOSC \citep[Li{\`e}ge Oscillation Code][]{LOSC}. One grid was employed for the modelling of the eclipsing binaries as well as NGC~6819. The second grid was constructed to model the stars in NGC~6791.

In the two grids, we vary both the initial mass and the metallicity, hereby exploring a two-dimensional parameter space. For the first grid, the initial helium abundance ($Y_\mathrm{i}$) is assumed to be related to the initial abundance of heavy elements ($Z_\mathrm{i}$), in such a way as that an increase in $Z_\mathrm{i}$ is accompanied by an equal increase in $Y_\mathrm{i}$, i.e. $\Delta Z_\mathrm{i}/\Delta Y_\mathrm{i} = 1.0$. As mentioned above, this grid is applied to model the eclipsing binaries and NGC~6819. For the second grid, we set $\Delta Z_\mathrm{i}/\Delta Y_\mathrm{i}$ to $2.0$, in order to recover the chemical properties of NGC~6791 \citep{Brogaard2012}. {\color{black}For all models presented in this paper, we do not consider alpha enrichment, i.e. $\mathrm{[\alpha/Fe]}=0.0$ (cf. Section~\ref{sec:NGC6791} for a discussion hereon).}

Both grids cover the evolution from the pre-main sequence to the red giant branch. For the first grid, we have computed stellar models with masses between $0.7$ and $2.5\,\mathrm{M}_\odot$ in steps of $0.02\,\mathrm{M}_\odot$. The grid includes 23 different values of $\mathrm{[Fe/H]}$, ranging from -2.5 to $0.2\,$dex. As regards the metallicity, the step-size is not uniform but alters between 0.10 and $0.15\,$dex. The grid contains stellar models with radii up until $25\,\mathrm{R}_\odot$.

For the second grid, with which we address NGC~6791, we have again computed stellar models with masses between $0.7$ and $2.5\,\mathrm{M}_\odot$  in steps of $0.02\,\mathrm{M}_\odot$. The grid includes models with metalicities between -0.1 to $0.5\,$dex in steps of $0.05\,$dex.

For both grids, the composition of the models is based on the solar mixture evaluated by \citep{Asplund2009}. We use the FreeEOS by A.~W. Irwin \citep{Cassisi2003}, the nuclear reaction rates by \cite{Adelberger2011}, and the semi-empirical $T(\tau)$ relation by \cite{1981ApJS...45..635V}.
We employ OPAL opacities \citep{Iglesias1996} in combination with the low-temperature opacities by \cite{Ferguson2005}. We have included both over- and undershooting, setting the associated parameters {\color{black}($\alpha_\mathrm{ov}$ and $\alpha_\mathrm{un}$)} to 0.1 and 0.2 for over- and undershooting, respectively. \textsc{cl{\'e}s} uses instantaneous overshooting. {\color{black}The extent of the over- and undershooting regions are $\alpha_\mathrm{ov} H_\mathrm{P}$ and $\alpha_\mathrm{un} H_\mathrm{P}$, respectively, where $H_\mathrm{P}$ denotes the pressure scale height. In the case of convective core overshooting, we substitute $H_\mathrm{P}$ by the size of the convective core ($r_\mathrm{cc}$), if $H_\mathrm{P}>r_\mathrm{cc}$. In the over- and undershooting regions, we use the radiative temperature gradient.}

\subsection{Likelihood} \label{sec:likelihood}

When comparing models to data, we include spectroscopic constraints on the effective temperature ($T_\mathrm{eff}$) as well as on the metallicity ($\mathrm{[Fe/H]}$), assuming that these measurements are uncorrelated and that the noise is Gaussian. 

In addition, we include the individual radial mode frequencies ($\ell=0$) as asteroseismic constraints --- without directly imposing constraints on the radial order of each mode. The individual frequencies are compared to observations after applying a surface correction specified in Section~\ref{sec:surfcorr}. While adding non-radial mode frequencies may help to further constrain the stellar parameters, we limit ourselves to radial modes in this analysis. The reason for this choice is that we perform a differential study, in which we seek to compare the different surface correction relations on equal footings. Thus, while the surface correction relation by \cite{Ball2014} applies to all modes, the surface correction relations by both \cite{Kjeldsen2008} and \cite{Sonoi2015} are only valid for radial modes\footnote{This being said, as regards the power-law correction relation by \cite{Kjeldsen2008}, some authors have included non-radial modes by addressing the effect of the mode inertia \citep[e.g.][]{2013Sci...342..331H}.}. 
We are, in other words, restricted in our choices regarding the seismic constraints by the prescriptions we seek to compare.
As in the case of the spectroscopic constraints, the noise of each radial mode frequency is assumed to be Gaussian and uncorrelated with the remaining observed frequencies. This is a reasonable approximation for the considered modes. 

Regarding priors, we only include the constraints on the parameters that enter the surface correction relations specified in Section~\ref{tab:param}. In other words, we do not include any further prior restrictions on the global stellar properties --- other than those that are indirectly imposed through the limited extent of our grids.

Following the described approach, however, we often find the posterior distributions of different stellar parameters to be multi-modal. Furthermore, for one of the explored surface correction relations (see Section~\ref{sec:EB}), the associated {\'e}chelle diagrams show that the inferred surface effect often exceeds the observed large frequency separation ($\Delta \nu$) and is a significant fraction of the frequency of maximum power ($\nu_\mathrm{max}$). In other words, for these cases, the free parameters that enter the surface correction relation are chosen by \textsc{aims}, in such a way as to shift the model frequencies substantially. We deem such solutions to be un-physical. We will discuss these findings further in Sections~\ref{sec:EB} and \ref{sec:NGC6819}.

One viable approach to address both issues mentioned above would be to make the priors on the global stellar parameters more informative. Alternatively, one may make the likelihood more informative. We settled for the latter approach, including $\nu_\mathrm{max}$ into the likelihood, using the scaling relation by \cite{1991ApJ...368..599B} and \cite{Kjeldsen1995}:
\begin{equation}
\nu_\mathrm{max} = \left(\frac{M}{\mathrm{M}_\mathrm{\odot}}\right)\left(\frac{R}{\mathrm{R}_\mathrm{\odot}}\right)^{-2}\left(\frac{T_\mathrm{eff}}{\mathrm{T}_\mathrm{eff\odot}}\right)^{-1/2}\nu_\mathrm{max\odot}. \label{eq:numax} 
\end{equation}
{Here, $M$ and $R$ denote the stellar mass and photospheric radius, respectively.

The individual stellar acoustic mode frequencies tightly constrain the mean density. By including $\nu_\mathrm{max}$ into the likelihood, we introduce constraints on $\log g$ and thus lift degeneracies between the inferred stellar mass and radius, which explains the improved inference.}
Studies by \cite{Handberg2017} and \cite{Zinn2019} show that Eq.~(\ref{eq:numax}) is both accurate and precise for stars on the RGB with the considered metallicities \citep[see also ][]{Viani2017} --- \cite{Zinn2019} show that this statement holds true to a $2\,\%$ based on Gaia parallaxes \citep{Gaia}. Below, we discuss and compare the results obtained with and without constraints on $\nu_\mathrm{max}$ to determine the implications of adding this quantity into the likelihood.

{The fact that we obtain more accurate results by including $\nu_\mathrm{max}$ goes to show the importance of additional constraints, including non-seismic measurements, that complement the individual observed frequencies.}
We note that we could have introduced other (non-seismic) constraints into the likelihood --- of course, each of these come with their own caveats. One option would be to use the stellar luminosity. However, adding such constraints or exploring the ramifications of doing so would not contribute to answering the scientific question that we address in this paper: how does the use of specific surface correction relations affect the inferred stellar properties. Indeed, the use of additional or alternative non-seismic constraints might obscure the influence of the surface correction relations by dominating the likelihood. {In other words, while good non-seismic constraints are often invaluable for asteroseismic analyses, they do not help us to discriminate between the different surface correction relations.} As stated above, we, therefore, restrict ourselves to including weak spectroscopic constraints on $T_\mathrm{eff}$ and $\mathrm{[Fe/H]}$ in addition to the constraints on $\nu_\mathrm{max}$.

{
\subsection{Goodness of fit}

Throughout this paper, we discuss the goodness of fit for the maximum a posteriori models, i.e. the best-fitting models, within each run. As a measure for the goodness of fit, we refer to the reduced $\chi^2$-value of the best-fitting models:
\begin{equation}
\chi^2_\mathrm{red} = \frac{1}{N} \sum_{i=1}^{N} \frac{(x_{\mathrm{mod},i}-x_{\mathrm{obs},i})^2}{\sigma_i^2}. \label{eq:chi2}   
\end{equation}
Here, the sum runs over all $N$ observational constraints, and the subscripts 'mod' and 'obs´ refer to the model and observational values, respectively. The models, for $\chi^2_\mathrm{red}\approx 1$, reliably recover the observational constraints. While models that yield $\chi^2_\mathrm{red} \gg 1$ simply constitute a poor fit, values of $\chi^2_\mathrm{red}\ll 1$ imply that the data is overfitted.

As an alternative measure for the goodness of fit, we directly refer to the evaluated likelihood function, from which we compute the so-called Bayesian information criterion (BIC):
\begin{equation}
\mathrm{BIC} = \ln(N_\mathrm{s})k-2\ln(\hat{\mathcal{L}}), \label{eq:BIC}   
\end{equation}
where $N_\mathrm{s}$ denotes the combined number of samples from the Markov chains, $k$ is the number of free parameters, and $\hat{\mathcal{L}}$ denotes the maximized value of the likelihood function. Better models lead to lower values of the BIC: the second term in Eq.~(\ref{eq:BIC}) rewards approaches, i.e. combinations of stellar models and surface correction relations, that yield high values for the maximized likelihood. The first term in Eq.~(\ref{eq:BIC}), meanwhile, penalizes models that reach this goal due to a high degree of complexity.
}

\section{surface correction relations} \label{sec:surfcorr}

The surface effect describes a systematic offset ($\delta \nu$) between {\color{black}the uncorrected adiabatic} model frequencies ($\nu_\mathrm{mod}$) and observations ($\nu_\mathrm{obs}$):
\begin{equation}
\delta \nu = \nu_\mathrm{obs} - \nu_\mathrm{mod}.
\end{equation}
As discussed in the introduction, this frequency offset comes about as a result of the structural inadequacies of state-of-the-art stellar models as well as simplifying approximations that enter the frequency computation.

{\color{black}When determining the likelihood of each model in the Markov chain by comparing model frequencies to observations, we correct the model frequencies, taking the surface effect into account. We do so using different surface correction relations. We discuss each of these surface correction relations below.
}

To mitigate the surface effect, \cite{Kjeldsen2008} propose to fit the surface effect by a power law, calibrating the exponent ($b$) based on the present-day Sun:
\begin{equation}
\frac{\delta \nu}{\nu_\mathrm{max}} =  a \left(\frac{\nu_\mathrm{obs}}{\nu_\mathrm{max}} \right)^b \label{eq:poly}.
\end{equation}
Throughout the rest of this paper, we refer to the surface correction relation in Eq.~(\ref{eq:poly}) as K08. We note that \textsc{aims} uses $\nu_\mathrm{mod}$ rather than $\nu_\mathrm{obs}$ when adopting Eq.~(\ref{eq:poly}) under the assumption that $\delta \nu \ll \nu_\mathrm{obs}$.

When following the approach suggested by \cite{Kjeldsen2008}, $b$ is commonly set to the solar calibrated value of 4.9 \citep[e.g.][]{Nsamba2018}, while $a$ is adjusted based on the observed frequencies. Drawing upon the analysis of 3D simulations by \cite{Sonoi2015}, one may alternatively vary both $a$ and $b$ as a function of the global stellar parameters:
\begin{equation}
\log |a| = 8.13 \log T_\mathrm{eff} - 0.670 \log g - 30.2, \label{eq:loga}
\end{equation}
and $a$ is negative, while
\begin{equation}
\log b = - 3.16 \log T_\mathrm{eff} + 0.184 \log g + 11.7. \label{eq:logb}
\end{equation}

\cite{Sonoi2015} argue in favour of substituting the power-law fit by a Lorentzian surface correction relation, as this yields a better fit to the frequency shift derived from 3D MHD simulations within the gas $\Gamma_1$ approximation mentioned above:
\begin{equation}
\frac{\delta \nu}{\nu_\mathrm{max}} =  \alpha \left(1-\frac{1}{1-(\nu_\mathrm{obs}/\nu_\mathrm{max})^{\beta}} \right). \label{eq:lorentz}
\end{equation}
In the following, we refer to the surface correction relation in Eq.~(\ref{eq:lorentz}) as S15. According to the analyses by \cite{Sonoi2015}
\begin{equation}
\log |\alpha| = - 7.69 \log T_\mathrm{eff} - 0.629 \log g -28.5, \label{eq:logalpha}
\end{equation}
and $\alpha$ is negative, while
\begin{equation}
\log \beta = - 3.86 \log T_\mathrm{eff} + 0.235 \log g + 14.2. \label{eq:logbeta}
\end{equation}
As in the case of Eq.~(\ref{eq:poly}), \textsc{aims} draws on $\nu_\mathrm{mod}$ rather than $\nu_\mathrm{obs}$ when correcting model frequencies.

Finally, \cite{Ball2014} presents a surface correction relation based on an asymptotic analysis by \cite{1990LNP...367..283G}:
\begin{equation}
\frac{\delta \nu_i}{\nu_\mathrm{ac}} = \mathcal{I}^{-1} \left( a_{-1}\frac{ \nu_{\mathrm{ac}}}{\nu_\mathrm{mod}} + a_3 \frac{\nu_\mathrm{mod}^3}{\nu_{\mathrm{ac}}^3} \right) \label{eq:ball}.
\end{equation}
Here, $a_{-1}$ and $a_3$ are free parameters, $\nu_{\mathrm{ac}}$ denotes the acoustic cutoff frequency that scales linearly with $\nu_\mathrm{max}$, and $\mathcal{I}$ is the mode inertia. We note that we substitute $\nu_{\mathrm{ac}}$ by a reference frequency that is related to the dynamical time-scale in \textsc{aims}.
In the following, we refer to the surface correction relation in Eq.~(\ref{eq:ball}) as BG14.

When faced with free parameters in any of the surface correction relations mentioned above, \textsc{aims} selects that combination of parameters that minimizes the discrepancy between model frequencies and observations. This is done by following the procedure suggested by \cite{Ball2014}. In other words, when these parameters ($a$, $b$, $\alpha$, $\beta$, $a_{-1}$, or $a_3$) are kept free, they are optimized for every single stellar model rather than being randomly sampled by the MCMC. In this way, the free parameters of the surface correction relations adjust based on the pseudo-random walk through the parameters space spanned by the stellar mass and composition. 

We note that the free parameters that enter the surface correction relations depend on both $\nu_\mathrm{mod}$ and $\nu_\mathrm{obs}$, which implies that the corrected model frequencies are correlated both internally and with $\nu_\mathrm{obs}$. Following common practise, we do not take these correlations into account.

The following section gives an overview of which parameters are kept fixed and which parameters are varied in different approaches.

\subsection{Choosing surface correction relation parameters}

In this paper, we deal with nine distinct ways of addressing the surface effect. We hence both employ different surface correction relations and vary the prescriptions for the different parameters involved. A summary can be found in Table~\ref{tab:param}.

\begin{table}
    \centering
    \caption{Summary of the different surface correction relations that are investigated in this paper.}
    \label{tab:param}
    \begin{tabular}{lccccccccccccccccccccc} 
        \hline
        Surf. corr. & Eq. & Parameters \\
        \hline
        K08 (a) & \ref{eq:poly} & $ a \leq 0$, $b=4.9$ \\ [2pt]
        K08 (b)  & \ref{eq:poly} & $ a\leq 0$, $b$ from Eq.~(\ref{eq:logb})\\ [2pt]
        K08 (c) & \ref{eq:poly} & $ a \leq 0$, $b \geq 0$ \\ [2pt]
        S15 (a) & \ref{eq:lorentz} & $\alpha \leq 0$, $\beta =4.0$ \\ [2pt]
        S15 (b) & \ref{eq:lorentz} & $\alpha \leq 0$, $\beta$ from Eq.~(\ref{eq:logbeta}) \\ [2pt]
        S15 (c) & \ref{eq:lorentz} & $ \alpha \leq 0$, $\beta \geq 0$ \\ [2pt]
        BG14 (a) & \ref{eq:ball} & $a_{-1}=0$, $a_3$ is free \\ [2pt]
        BG14 (b) & \ref{eq:ball} & $a_{-1}$ and $a_3$ are free \\ [2pt] 
        NoSC (None) & -- & -- \\ [2pt]
        \hline
\end{tabular}
\end{table}

When using K08, we either fix $b$ to the solar calibrated value of 4.9 as found by \cite{Kjeldsen2008}, establish $b$ using Eq.~(\ref{eq:logb}), or let $b$ adjust freely. In the latter case, we require that $b\geq 0$, since the surface effect increases with increasing frequency. This is due to the frequency dependence of the {\color{black}upper} turning point of the oscillation. {\color{black}High-frequency modes thus probe shallower near-surface layers than low-frequency modes do, i.e. the eigenfunctions of low-frequency modes are evanescent in the near-surface layers \citep[cf.][for a detailed discussion]{1997MNRAS.284..527C}.} In all three cases, we let $a$ vary freely, requiring that $a\leq 0$. This requirement builds on the assumption that the combined surface effect is negative as in the case of the present-day Sun and other main-sequence stars \citep[e.g.][]{Brown1984,Christensen-Dalsgaard1988,Houdek2017,Houdek2019}. 

When dealing with S15, we similarly distinguish between three different approaches: in the first approach, we fix $\beta$ to 4.0 as done by \cite{Nsamba2018}. In the second approach, $\beta$ is determined using Eq.~(\ref{eq:logbeta}). Finally, we let $\beta$ vary, solely requiring that $\beta \geq 0$. In all cases, we let $\alpha$ adjust freely, requiring that $\alpha \leq 0$, following the arguments given above.

{\color{black}As regards the cases of K08 and S15, for which we let both parameters adjust freely, we stress that this path is not commonly taken in literature. This is because the sole physical justification for K08 and S15 lies in the calibrated parameters. Indeed, as shown below, K08 and S15 do not perform well, when no such calibration has taken place. By allowing both parameters to adjust freely, we thus are able to highlight the limitations of K08 and S15.}

We furthermore investigate two different cases based on BG14: in one case, we only include the cubic term, while we allow both coefficients ($a_{-1}$ and $a_3$) to vary freely in the second approach. In other words, we include both a one- and a two-term version of BG14.

For comparison, we include the case, where no surface correction is taken into account. In the following, we refer to this as NoSC.

\section{Eclipsing Binaries} \label{sec:EB}

Using \textsc{aims}, we have determined the stellar parameters of eight red giant branch stars in the \textit{Kepler} field: 
{KIC~4054905, KIC~4663623}, KIC~5786154, KIC~7037405, KIC~8410637, KIC~8430105, {KIC~9540226}, and KIC~9970396. All eight stars are members of eclipsing {and spectroscopic} binaries, which allows for accurate dynamical measurements of their masses and radii. 
This property makes the investigated eight giants suitable benchmark stars for asteroseismic analyses \citep[e.g.][]{Gaulme2016,Tanda2018}. We thus compare our results with the conclusions from dynamical studies by \cite{Gaulme2016}, \cite{Brogaard2018}, and Benbakoura et al. (submitted). The observational constraints are summarized in Table~\ref{tab:const1}.

\begin{table}
    \centering
    \caption{Summary of the observational constraints that were employed to model the eclipsing binaries investigated in this paper. This includes the number of radial modes to which we refer as $\Delta n$. The observed frequencies were deduced using PBjam or using the method presented by Gaulme et al. (2009). The spectroscopic constraints on $T_\mathrm{eff}$, $\mathrm{[Fe/H]}$, and $\nu_\mathrm{max}$ stem from Gaulme et al. (2016), Brogaard et al. (2018), Li et al. (2018), and Benbakoura et al. (submitted).}
    \label{tab:const1}
    \begin{tabular}{lccccccccccccccccccccc} 
        \hline
         & $\Delta n$ & $T_\mathrm{eff}$ [K] & $\mathrm{[Fe/H]}$ & $\nu_\mathrm{max}$ \\
        \hline
        KIC~4054905 & 6 & $ 4790 \pm 190 $  & $ -0.72 \pm 0.31 $ & $ 48.15 \pm 0.21 $ \\ [2pt]
        KIC~4663623 & 7 & $ 4803\pm 91 $ & $0.16 \pm 0.04$ & $ 46.51 \pm 2.34 $ \\ [2pt]
        KIC~5786154 & 5 & $4747 \pm 100$ & $-0.06 \pm 0.06$ & $29.75 \pm 0.16$ \\ [2pt]
        KIC~7037405 & 5 & $4500 \pm 80$ & $-0.27 \pm 0.05$ & $21.75 \pm 0.14$ \\ [2pt]
        KIC~8410637 & 7 & $4699 \pm 91$ & $0.16 \pm 0.05$ & $46.00 \pm 0.19$ \\ [2pt]
        KIC~8430105 & 8 & $5042 \pm 68$ & $-0.49 \pm 0.04$ & $76.70 \pm 0.57$ \\ [2pt]
        KIC~9540226 & 6 & $4662 \pm 91 $  & $ -0.16 \pm 0.08 $ & $ 27.88 \pm 0.17 $ \\ [2pt]
        KIC~9970396 & 6 & $4860 \pm 80$ & $-0.35 \pm 0.1$ & $63.70 \pm 0.16$ \\ [2pt]
        \hline
\end{tabular}
\end{table}

\begin{figure}
\centering
\includegraphics[width=1.00\linewidth]{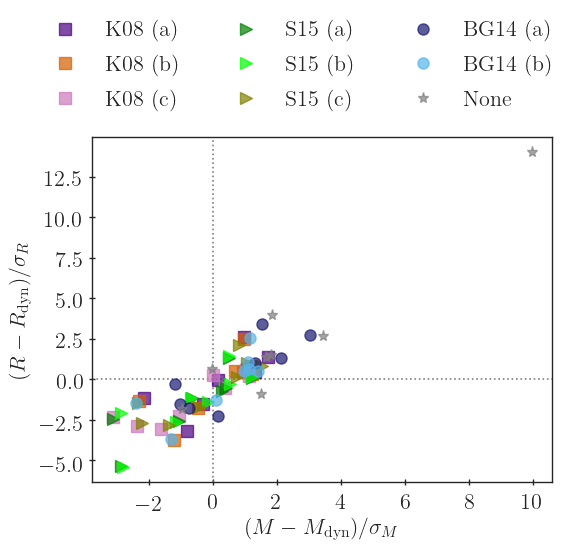}
\caption{{\color{black}Absolute} differences between dynamically inferred stellar parameters and the results obtained from stellar models when employing different surface correction relations. The differences are given in units of the standard deviation of the differences, i.e. $\sigma_M$ and $\sigma_R$ include both the observational errors and the uncertainty of the inferred model properties. The outlier, for which deviations in both mass and radius exceed 10 $\sigma$, is associated with KIC~4054905.}
\label{fig:MRbina_diff}
\end{figure}

\begin{figure}
\centering
\includegraphics[width=1.00\linewidth]{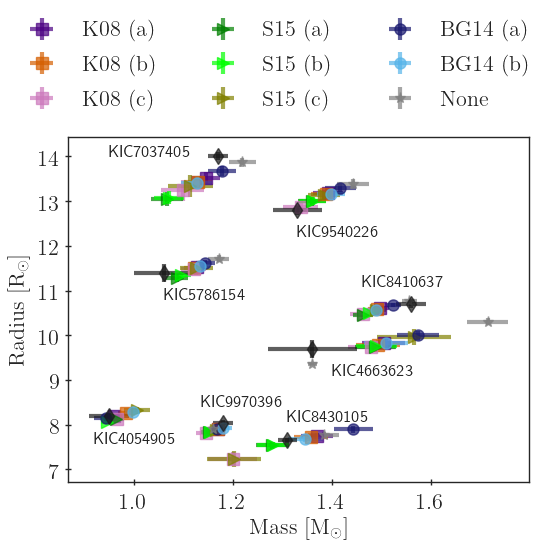}
\caption{Comparison between dynamically inferred stellar parameters (black markers) and the results obtained from stellar models when employing different surface correction relations. The markers show the location of the best-fitting model, while the error bars correspond to $68\,\%$ credibility intervals. The likelihood includes $\nu_\mathrm{max}$. When repeating the analysis using the medians rather than the best-fitting values, we reach the same quantitative and qualitative conclusions. Note that NoSC dramatically overestimates the mass and radius of KIC~4054905, yielding values that lie closer to KIC~4662623. This is the outlier in Fig.~\ref{fig:MRbina_diff}.}
\label{fig:MRbina}
\end{figure}

For all investigated nine treatments of the surface effect, the best-fitting model, i.e. the model with highest posterior probability, as well as the median of the obtained posterior probability distributions recover the correct masses and radii of most or all stars within $3\,\sigma$. This includes NoSC, i.e. the case, where no surface correction relation has been implemented. KIC~4054905 constitutes a prominent exception: the attempt to model this star without any surface correction relation fails to recover the correct mass and radius within $10\,\sigma$, while all surface correction relations perform well. These results are illustrated in Figs \ref{fig:MRbina_diff} and \ref{fig:MRbina}. Here, $\sigma$ includes {\color{black}the errors on the model parameters that are inferred from the MCMC as well as the errors on the dynamically inferred parameters. In other words, $\sigma$ denotes the combination of errors that can be derived from the law of propagation of errors.}

Furthermore, we find that S15 systematically yields lower mass and radius estimates than both K08 and BG14 do. NoSC and the one-term correction by BG14 (case a) systematically yield higher mass and radius estimates than the other approaches. These systematic trends are also found for the clusters and thus reappear in Sections~\ref{sec:NGC6819} and \ref{sec:NGC6791}. Indeed, the analyses of the two clusters show the same relative performance of the different surface corrections relations.

{There is thus a scatter in the mass and radius estimates that are obtained based on the different surface correction relations. The lowest scatter is found for {\color{black}KIC~9970396}. On first glance, Fig.~\ref{fig:MRbina} might suggest that a similarly low scatter is obtained for KIC~4054905. However, in reality, NoSC overestimates the mass by more than $10\,\sigma$, as mentioned above, yielding mass and radius estimates that more closely matches the dynamical constraints on KIC~4663623 than those on KIC~4054905.}

We partly ascribe the fact that all nine treatments of the surface effect perform similarly well to the included spectroscopic constraints and the constraints on $\nu_\mathrm{max}$. Thus, when no constraints on $\nu_\mathrm{max}$ are taken into account, a slightly different picture emerges: when Eq.~(\ref{eq:numax}) is not imposed, NoSC performs poorly, overestimating the radius obtained by \cite{Gaulme2016} and \cite{Brogaard2018} by up to a factor of two. The resulting errors in the inferred masses or radii exceed $5\,\sigma$ in six out of eight cases --- reaching $60\,\sigma$ for the radius of KIC~9970396. Moreover, only the two-term correction by BG14 (case b) does not have one or more outliers, for which the error in either the mass or radius exceeds $3\,\sigma$. Indeed, the two-term correction by BG14 (case b) recover the dynamical constraints on all binaries within $2.4\,\sigma$. For most of the remaining surface correction relations\footnote{In this particular case, NoSC does very well, recovering the dynamical constraints within $1\,\sigma$.}, KIC~8430105 is one of the outliers. However, we note that the power spectrum for KIC~8430105 is noisy and that the mode identification might be skewed due to the influence of binarity and magnetic activity {\citep[cf.][for further discussions on the influence of magnetic activity on stellar oscillations]{Magic2016,Perez2019}.}

In addition, when excluding $\nu_\mathrm{max}$ from the constraints and freely adjusting both parameters related to K08 (case c), the inferred surface effect is larger than the observed large frequency separation for KIC~8410637, KIC~8430105, and KIC~9540226. The surface effect is even a substantial fraction of the observed $\nu_\mathrm{max}$. Both the uncorrected theoretical values for $\Delta \nu$ and $\nu_\mathrm{max}$ are substantially larger than the observed values. This highly un-physical behaviour of K08 (case c) is likewise obtained for seven out of the nineteen stars that enter the analyses of NGC~6819 in Section~\ref{sec:NGC6819}. We attribute this behaviour to the fact that the physical justification for K08 solely lies in the calibration of the involved parameters. Without sufficient constraints on $a$ and $b$ in Eq.~(\ref{eq:poly}), a power-law description of the surface effect becomes unreliable. 

As shown by \cite{Ball2017} and \cite{Nsamba2018}, the use of solar calibrated values in K08 systematically shifts the stellar parameter estimates when addressing other main-sequence stars. Moreover, calibrating the involved parameters based on 3D MHD simulations as suggested by \cite{Sonoi2015} (Eq.~\ref{eq:logb}) has several caveats: firstly, current calibrations suffer from the use of the gas $\Gamma_1$ approximation, as discussed in the introduction. Secondly, the calibration by \cite{Sonoi2015} does not take deviations from solar metallicity into account \citep[cf.][who address this issue using 3D MHD simulations]{Manchon2018,Joergensen2019}. These points generally discourage from the use of K08 and also apply to S15. Nevertheless, disregarding the case of KIC~8430105, all surface correction relations explored in this paper perform equally well in recovering the global stellar mass and radius of the considered eclipsing binaries. Neglecting the surface effect altogether by not including a surface correction relation, however, does not yield the correct stellar parameters. {Indeed, we note that NoSC generally tends to overestimate the stellar mass and radius of the eclipsing binaries, while we have not spotted similarly clear trends for the remaining treatments of the surface effect. To shed more light on this issue, additional data points, i.e. more eclipsing binaries, are needed.}

\begin{figure}
\centering
\includegraphics[width=1.00\linewidth]{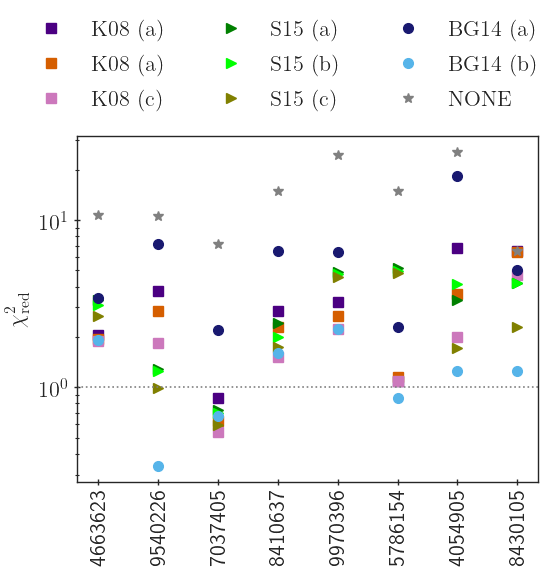}
\caption{Reduced $\chi^2$-values for all eight investigated eclipsing binaries (cf. Eq.~\ref{eq:chi2}). The labels on the abscissa indicate respective the KIC-numbers. The likelihood includes $\nu_\mathrm{max}$.}
\label{fig:chi2_bina}
\end{figure}

{Regarding the goodness of fit, we find several consistent trends. The best-fitting models that were selected using the one-term correction by BG14 (case a) and NoSC lead to higher reduced $\chi^2$-values and BICs than any of the other treatments of the surface effect do. For all eight stars, the best-fitting model with the lowest reduced $\chi^2$- and BIC-value is found by either using the two-term correction by BG14 (case b) or the free fit based on K08 (case c) --- the associated scatter is higher for case c of K08. For all investigated binaries, the two-term correction by BG14 (case b) thus {\color{black}results in good fits to data and often yields better fits than the remaining surface correction relations do in terms of the reduced $\chi^2$- and BIC-value}. We illustrate this in Fig.~\ref{fig:chi2_bina}
}

\begin{figure}
\centering
\includegraphics[width=1.00\linewidth]{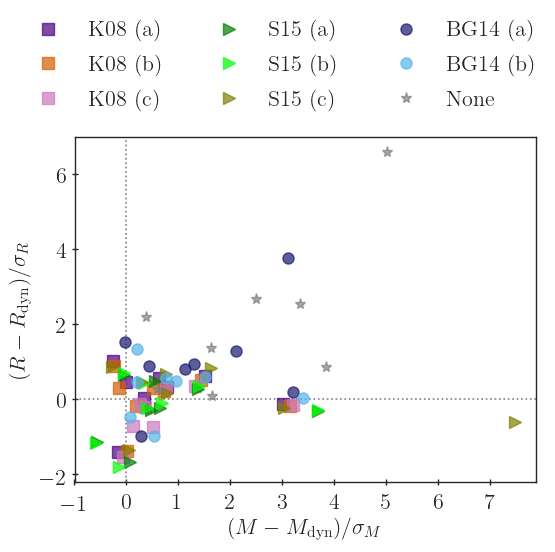}
\caption{As Fig.~\ref{fig:MRbina_diff} but employing the stellar radius rather than $\nu_\mathrm{max}$ in the likelihood. The accumulation of models, for which $(M-M_\mathrm{dyn})/\sigma_M\approx3$ and $(R-R_\mathrm{dyn})/\sigma_R\approx0$, is associated with KIC~7037405, for which we use the dynamical constraints by Brogaard et al. (2018). For these models, much lower residuals are obtained when using the dynamical constraints by Gaulme et al. (2016).}
\label{fig:MRbinaR_diff}
\end{figure}

\begin{figure}
\centering
\includegraphics[width=1.00\linewidth]{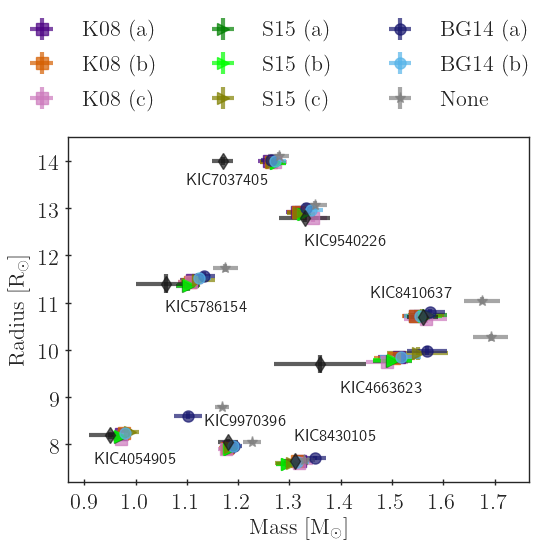}
\caption{As Fig.~\ref{fig:MRbina} but employing the stellar radius rather than $\nu_\mathrm{max}$ in the likelihood. Both the one-term correction of BG14 (case a) and NoSC dramatically overestimates the mass KIC~4054905, yielding a value that lies closer to KIC~9970396.}
\label{fig:MRbinaR}
\end{figure}

Since we are dealing with eclipsing binaries, accurate and precise alternative observational constraints are available. Among other parameters, the stellar radii have been well-determined. We have thus repeated the analysis, replacing the constraint on $\nu_\mathrm{max}$ with a constraint on the stellar radius. We find that this greatly reduces the scatter between the mass estimates obtained from the use of different surface correction relations. By including the radius into the observation constraints, we also reduce the mean absolute error in the inferred properties for all treatments of the surface effect.
We illustrate this in Figs~\ref{fig:MRbinaR_diff} and \ref{fig:MRbinaR}. Once again, this goes to show the importance of additional (non-seismic) constraints in asteroseismic analyses.
{The demonstrated improvement that arises from including the radius into the likelihood is, furthermore, of particular interest, since the upcoming third data release (DR3) by Gaia \citep{Gaia} will provide the community with robust measurements of stellar radii.} We note, however, that NoSC does still not perform as well as any of the surface correction relations do, even if we include the radius among the constraints. This underlines the vital role played by the treatment of the surface effect.

As mentioned above, the employed constraints on the stellar radius reduce the scatter between the inferred masses. To this end, the constraints on the stellar radius do not help us much to further discriminate between the different surface correction relations. On the other hand, this result goes to show that we can easily construct a likelihood that is not dominated by the bias introduced by the employed surface correction relation. In this connection, it is worth highlighting the consequences for the inferred properties of KIC~7037405. With the likelihood that includes the stellar radius, all surface correction relations unanimously point towards a mass of KIC~7037405 that is in better agreement with the dynamical measurements by \cite{Gaulme2016} ($1.25\pm 0.03\,\mathrm{M}_\odot$) than with the constraints by \cite{Brogaard2018} ($1.17 \pm 0.02 \mathrm{M}_\odot$), despite the fact that we use the constraints on the radius from \cite{Brogaard2018}. We note that this finding is furthermore in agreement with other asteroseismic studies \citep[cf. Fig.~14 in][]{2019MNRAS.482.2305B}.
{\color{black}However, we also note that the discrepancy between our models and the dynamical measurements by \cite{Brogaard2018} might partly come down to the chosen input physics. The models, for instance, follow a fixed relationship between the metal abundance and the helium content (cf. Section~\ref{sec:grid}), which might introduce a bias. However, we are partly able to delve into this issue, since KIC~7037405 also lies within the grid that was computed for the analysis of NGC~6791 in Section~\ref{sec:NGC6791}. As mentioned above, this grid employs a different value for $\Delta Z_\mathrm{i}/\Delta Y_\mathrm{i}$. We have thus repeated the analysis using another relationship between $Z_\mathrm{i}$ and $Y_\mathrm{i}$. We find that we arrive at the same conclusion: our results prefer the dynamical constraints by \cite{Gaulme2016} over those of \cite{Brogaard2018}. Moreover, we have repeated the analysis, altering the constraints on $\mathrm{[Fe/H]}$. This likewise leaves the conclusion unaffected.
}

In so far as that we believe that our models recover the correct stellar structures, our analysis of KIC~7037405 exemplifies how an asteroseismic analysis is able to complement the classical methods, irrespectively of the employed surface correction relation. {On the other hand, we note that surface correction relations do not account for all physical inadequacies of stellar models but only deal with the surface effect and that asteroseismic analyses may naturally fail to deliver accurate stellar parameter estimates if the underlying stellar models are incorrect \citep[see e.g.][for a comparison of models that are based on different physical assumptions]{Joergensen2019,2019MNRAS.484..771R}.}

{It is worth mentioning that the best-fitting models, as well as the mean of the associated posterior distributions, tend to systematically underestimate the measured effective temperatures for all binaries. Thus, the mean absolute deviation between measurements and model predictions exceeds $100\,$K in all cases. This holds true, whether or not we introduce $\nu_\mathrm{max}$ or the radius as an additional observational constraint.

As regards the goodness of fit, we obtain slightly higher values for both the reduced $\chi^2$ and the BIC when including the radius in the likelihood than shown in Fig.~\ref{fig:chi2_bina}. While an independent constraint on the stellar radius thus improves the agreement of the predicted stellar mass with dynamical measurements, it leads to a slightly worse recovery of the individual frequencies than when the individual frequencies dominate the likelihood. 
}

To further discriminate between the different treatments of the surface effect, more samples are needed. We, therefore, turn to an analysis of the open clusters NGC~6819 and NGC~6791 below.

\section{NGC 6819}  \label{sec:NGC6819}

Using \textsc{aims} and employing all nine treatments of the surface effect listed in Table~\ref{tab:param}, we have derived stellar parameters for 19 RGB stars in the open cluster NGC~6819\footnote{KIC~4937576, KIC~5023732, KIC~5023931, KIC~5024240, KIC~5024297, KIC~5024312, KIC~5024405, KIC~5024512, KIC~5024583, KIC~5111718, KIC~5111940, KIC~5112072, KIC~5112361, KIC~5112734, KIC~5112744, KIC~5112880, KIC~5112948, KIC~5113041, and KIC~5113441.}. We have adopted non-seismic constraints from \cite{Handberg2017} and set $\mathrm{[Fe/H]=0.02\pm0.1}\,$dex for all cluster members.

\begin{figure}
\centering
\includegraphics[width=1.00\linewidth]{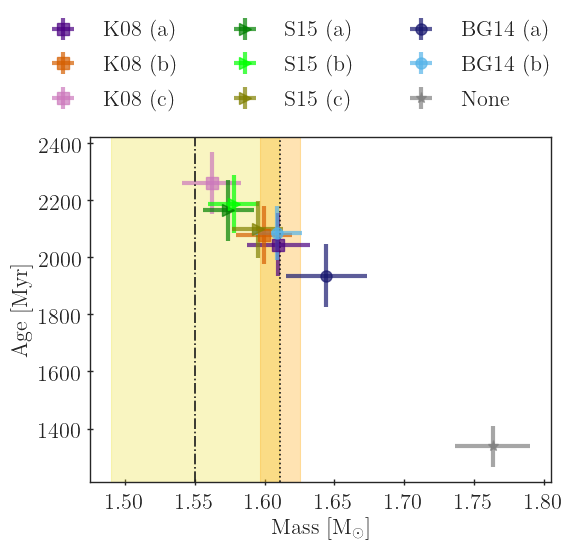}
\caption{The mean mass and age obtained from nine different treatments of the surface effect using Eqs (\ref{eq:meanM}) and (\ref{eq:uncert1}) {\color{black}for NGC~6819}. The orange shaded area indicates to the $68\,\%$ credibility interval of the asteroseismic constraints by Handberg et al. (2017). The corresponding mean is marked by the dotted black line. We note that the constraints obtained by Handberg et al. (2017) are likewise obtained from an asteroseismic study. Classical measurements by Sandquist et al. (2013) and Brogaard et al. (2015) lead to a broader confidence interval that agrees well with the results obtained based on all surface correction relations ($1.55\pm 0.06\,\mathrm{M}_\odot$). These constraints are shown with the shaded yellow area. The corresponding mean is indicated with the dash-dotted line. The likelihood includes $\nu_\mathrm{max}$.}
\label{fig:SumAM}
\end{figure}

We computed the mean mass of the 19 RGB stars based on each of the nine treatments of the surface effect. The results are summarized in Fig.~\ref{fig:SumAM} and in Table~\ref{tab:NGC6819}. When repeating the analysis without including $\nu_\mathrm{max}$ into the likelihood, all treatments lead to similar mean masses and mass scatter to those obtained with $\nu_\mathrm{max}$. We note that the presented averages are taken over the best-fitting values and weighted by the respective variance of the associated posterior probability distribution. We also note that we reach the same quantitative and qualitative conclusions if we repeat the analysis using the medians rather than the best-fitting values.
The applicability of this procedure rests on several assumptions, including the notion that the posterior probability distribution of the stellar mass is well-described by a Gaussian. While this is a good approximation for the vast majority of the stars presented in this paper, we note that cases exist, for which the posterior mass distribution is less symmetric or even multi-modal.

\begin{table}
    \centering
    \caption{Summary of the inferred mean mass and age ($\tau$) of the RGB stars in NGC~6819 based on the different surface correction relations that are investigated in this paper. The likelihood includes $\nu_\mathrm{max}$.}
    \label{tab:NGC6819}
    \begin{tabular}{lccccccccccccccccccccc} 
        \hline
        Surf. corr. & $\langle M \rangle \pm \sigma_{\bar{M}}$ & $\langle \tau \rangle \pm \sigma_{\bar{\tau}}$ [Myr] \\
        \hline
        K08 (a) & $1.610\pm 0.023$ & $2045\pm110$\\ [2pt]
        K08 (b) & $1.600\pm0.020$ & $2077\pm102$ \\ [2pt]
        K08 (c) & $1.562\pm0.021$ & $2260\pm108$ \\ [2pt]
        S15 (a) & $1.574\pm0.018$ & $2164\pm106$ \\ [2pt]
        S15 (b) & $1.578\pm0.0182$ & $2187\pm103$ \\ [2pt]
        S15 (c) & $1.595\pm0.018$ & $2098\pm98$ \\ [2pt]
        BG14 (a) & $1.644\pm0.029$ & $1935\pm110$ \\ [2pt]
        BG14 (b) & $1.609\pm0.018$ & $2084\pm94$ \\ [2pt] 
        NoSC (None) & $1.763\pm0.027$ & $1337\pm70$ \\ [2pt]
        \hline
\end{tabular}
\end{table}

Thus, the mean mass of the cluster is estimated to be
\begin{equation}
\bar{M} = \frac{\sum_i^N M_i/\sigma_i^2}{\sum_i^N {\color{black}1/}\sigma_i^2}, \label{eq:meanM}  
\end{equation}
where $M_i$ and $\sigma_i$ denote the best-fitting value of the mass and the corresponding estimate of the standard deviation for each of the $N$ samples. Due to the low number of samples, the uncertainty on the mean has been computed by following the procedure outlined by \cite{Chaplin1998} \citep[see also][]{Miglio2012}. 
{\color{black}When dealing with small sample sizes, the estimate of the standard deviation that follows from the sample might be biased and often underestimates the standard deviation of the underlying normally distributed population. To account for this, we} compute the uncertainty from the unbiased estimate of the variance from the sample and apply a correction factor $t[N-1]$ drawn from the Student's $t$ distribution with $N-1$ degrees of freedom:
\begin{equation}
\sigma_{\bar{M}} = t[N-1] \sqrt{\frac{\sum_i^N (M_i-\bar{M})/\sigma_i^2}{(N-1)\sum_i^N {\color{black}1/}\sigma_i^2}}. \label{eq:uncert1}
\end{equation}
In the present case, the correction factor, i.e. {\color{black}$t[N-1]$}, is of the order of unity, {\color{black}largely leaving the uncertainties unaltered}. We note that Eq.~(\ref{eq:uncert1}) yields larger uncertainties than what would be obtained from the law of propagation of error based on the uncertainties on the samples alone. This implies that the scatter of the samples are not dominated by random errors, which might imply that \textsc{aims} underestimates the true errors on the sampled stellar parameters --- at least, within the Gaussian approximation. This circumstance might reflect the fact that the grid is two-dimensional, as additional free parameters might lead to broader posterior distributions of the sampled properties. {\color{black}As specified in Section~\ref{sec:grid}}, we thus use a fixed mixing length parameter{, fixed parameters associated with over- and undershooting,} and a fixed relation between the initial helium abundance and the initial abundance of heavy elements.

Based on dynamical measurements of eclipsing binaries by \cite{Sandquist2013}, \cite{Brogaard2015} determine the mean mass of the RGB stars in the NGC~6819 to be $1.55 \pm 0.06\,\mathrm{M}_\odot$. Using asteroseismic scaling relations and empirical corrections for $\Delta \nu$ and $\nu_\mathrm{max}$, \cite{Handberg2017} finds the mean mass of RGB stars in the cluster to be $1.61 \pm 0.02\,\mathrm{M}_\odot$ \citep[see also e.g.][]{Miglio2012}. We note that only the models, for which no surface correction relation is taken into account, fail to fall within one standard deviation of the mean evaluated by either \cite{Brogaard2015} or \cite{Handberg2017}. This holds true with and without including $\nu_\mathrm{max}$ into the likelihood. 

This being said, such a direct comparison with the mean cluster masses of RGB stars from \cite{Brogaard2015} and \cite{Handberg2017} might be slightly skewed since our analysis only includes 19 stars. Our results may thus be subject to a selection bias. Instead, we note that \cite{Handberg2017} also supply mass estimates for the individual members of the cluster, yielding a weighted mean mass of $1.61\pm0.01\mathrm{M}_\odot$ for the considered 19 giants.
With the exception of NoSC, we find that all treatments of the surface effect yield parameter estimates that agree with this result within $2\,\sigma$ whether or not we include $\nu_\mathrm{max}$ in the likelihood. Moreover, on a star {\color{black}by} star basis, all models recover the observational constraints on the effective temperature within $100\,$K, irrespectively of the employed surface correction relation. Again, this statement does not apply to NoSC but does remain valid whether or not we include $\nu_\mathrm{max}$ in the likelihood. We furthermore note that we find no obvious trends between the inferred mass of the individual cluster members and the associated value of $\nu_\mathrm{max}$ when including the latter in the likelihood. This holds true for all nine treatments of the surface effect --- and it also holds true for the stars in NGC~6791. Moreover, we find the mass scatter to be seemingly uniform as a function of $\nu_\mathrm{max}$, i.e. along the RGB.

As can be seen from Fig.~\ref{fig:SumAM}, the different treatments of the surface effect that employ S15 tend to lead to lower average masses than the two-term surface correction relation by BG14 does. This is consistent with our analysis of the eclipsing binaries, where S15 is likewise found to underestimate the classical constraints on the stellar masses. According to a differential study by \cite{Nsamba2018}, this behaviour is not found for main-sequence stars, where S15 and BG14 on average yield equally robust mass and age estimates. {The fact that the cases a and b of S15 perform better for main-sequence stars than for more evolved stars presumably reflects the fact that the involved parameters are calibrated based on a study that primarily includes patched models of main-sequence stars. Thus, \citep{Sonoi2015} solely include one red giant branch star in their calibration sample.} 

As regards S15, we furthermore note that we have excluded KIC~5112880 from the sample when treating $\alpha$ and $\beta$ as free parameters (case c). This is because case c of S15 yields an age estimate for this star that exceeded the age of the Universe by a factor of two. Once again, this behaviour might reflect the fact that there is no physical justification for S15 beyond the calibrated surface correction relation parameters.

From Fig.~\ref{fig:SumAM}, it is also apparent that the discrepancies in the estimated mean stellar mass translate into substantial systematic offsets in the estimated mean age of the stars: excluding the case, where the surface effect is left unaccounted for, the choice for the treatment of the surface effect alone affects the estimated age of the cluster by up to $17\,\%$.  {As discussed in Section~\ref{sec:likelihood}, tighter non-seismic constraints might lift degeneracies and hereby partly mend but not eliminate this issue. The obtained systematic offsets} thus illustrate the vital role that a proper understanding of the surface effect, i.e. superadiabatic convection, plays for galactic archaeology.

\subsection{Distance modulus}

To investigate how the treatment of the surface effect alters other inferred physical properties of the cluster, we have computed the apparent distance modulus, $(m-M)_\mathrm{V}$, of NGC~6819 based on the prescription by \cite{Torres2010}. For this purpose, we use the V-band magnitudes by \cite{2014AJ....148...38M} in combination with the bolometric corrections by \cite{Casagrande2014}.

\begin{figure}
\centering
\includegraphics[width=1.00\linewidth]{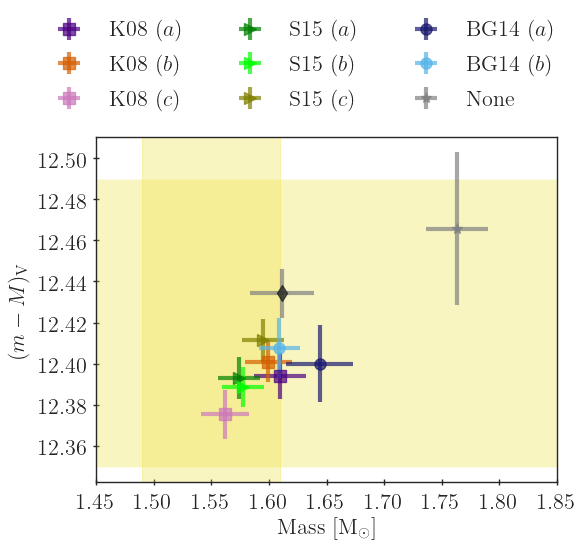}
\caption{The mean mass and distance modulus obtained from nine different treatments of the surface effect {\color{black}for NGC~6819}. The error bars on the distance modulus correspond to the sample standard deviation and thus reflect the scatter of the individual distance moduli. The black marker summarizes the results in Handberg et al. (2017) based on the same 19 stars. The yellow shaded area corresponds to the $68\,\%$ credibility interval for the classical constraints on the mass and distance modulus presented by Sandquist et al. (2013) and Brogaard et al. (2015). The likelihood includes $\nu_\mathrm{max}$.}
\label{fig:distance}
\end{figure}

In Fig.~\ref{fig:distance}, we compare the obtained results with classical measurements by \cite{Sandquist2013} and \cite{Brogaard2015}, according to whom the apparent distance modulus and mean mass of the RGB stars in the cluster are $12.42 \pm 0.07$ and $1.55\pm 0.06\,\mathrm{M}_\odot$, respectively. 
As can be seen from the figure, the influence of the treatment of the surface effect on the distance modulus is as pronounced as its influence on the age and mass estimates. Indeed, the spread of obtained mean apparent distance moduli are similar to the internal scatter.

We find that all surface correction relations recover the mean distance modulus of the cluster within $1\,\sigma$ whether or not we include $\nu_\mathrm{max}$ in the likelihood.

Based on the same 19 giants, \cite{Handberg2017} finds the apparent distance modulus of NGC~6819 to be $12.43 \pm 0.01$, which likewise agrees with the results presented by \cite{Sandquist2013} and \cite{Brogaard2015} within $1\,\sigma$. Notably, all the investigated treatments of the surface effect furthermore recover the quoted distance modulus by \cite{Handberg2017} within $2\,\sigma$ --- again this holds true with and without constraints on $\nu_\mathrm{max}$.

On a star {\color{black}by} star basis, we find the mass to be correlated with the attributed distance modulus. We thus find that the stars that have been assigned a lower mass are also assigned a lower distance modulus. We ascribe this behaviour to the correlation between seismically inferred masses and radii. The described correlation between the mass and distance modulus is observed with and without constraints on $\nu_\mathrm{max}$. We address this issue further in Section~\ref{sec:NGC6791}, where we deal with NGC~6791.


\section{NGC 6791} \label{sec:NGC6791}

In this section, we present an analysis of 30 red giants\footnote{KIC~2435987, KIC~2436688, KIC~2437270, KIC~2437402, KIC~2437972, KIC~2438140, KIC~2569618, KIC~2570244, KIC~2436097, KIC~2437653, KIC~2437933, KIC~2437976, KIC~2438333, KIC~2570094, KIC~2436540, KIC~2436818, KIC~2437240, KIC~2437488, KIC~2437957, KIC~2438038, KIC~2570172, KIC~2569360, KIC~2437816, KIC~2436824, KIC~2436814, KIC~2437444, KIC~2437507, KIC~2436900, KIC~2436209, KIC~2436332.} in the open cluster NGC~6791 to further validate and extend upon the conclusions drawn in Section~\ref{sec:NGC6819}. Based on \cite{Brogaard2012}, we set $\mathrm{[Fe/H]=0.29\pm0.1}\,$dex. Non-seismic constraints were adopted from \cite{Basu2011}.

\begin{figure}
\centering
\includegraphics[width=1.00\linewidth]{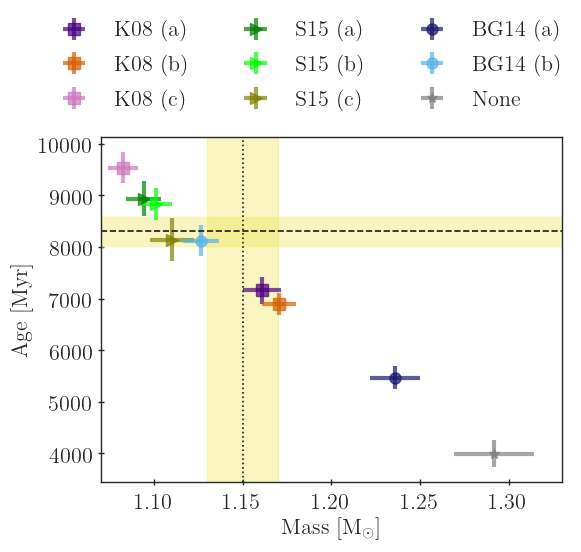}
\caption{As Fig.~\ref{fig:SumAM} but for NGC~6791. Here, we compare to the results listed in Brogaard et al. (2012). The horizontal dashed line and the yellow shaded horizontal area indicate an age estimate based on isochrones.}
\label{fig:SumAM_NGC6791}
\end{figure}

As for NGC~6819, we have computed the mean mass for the cluster based on the output from \textsc{aims} for all 30 considered members. We have repeated this for all nine surface correction relations in Table~\ref{tab:param}. The results are summarized in Fig.~\ref{fig:SumAM_NGC6791} as well as in Table~\ref{tab:NGC6791}. The figure includes an estimate for the absolute mass of stars on the lower red giant branch by \cite{Brogaard2012} ($1.15\pm 0.02 \,\mathrm{M_\odot}$) as well as an age estimate ($8.3\,\pm0.3$Gyr) based on isochrones from the same paper. While the results by \cite{Brogaard2012} are based on an analysis of eclipsing binaries, we note that the cited {\color{black}age estimate} is model dependent. As discussed in Section 3.2 of the paper by \cite{Brogaard2012}, the assumed abundances of different elements and the treatment of heavy element diffusion lead to additional systematic and statistical errors on the age estimate. Analogously to the case of NGC~6819, we furthermore note that the mass estimate from our analysis might be subject to a selection bias. After all, we only consider 30 stars. We also note that analyses by other authors yield slightly different mass estimates for the red giant branch stars in the cluster. This includes asteroseismic analyses by \cite{Basu2011} and \cite{Miglio2012} --- that is, $1.20\pm0.01\,\mathrm{M}_\odot$ and $1.23\pm0.02\,\mathrm{M}_\odot$, respectively. This being said, both \cite{Basu2011} and \cite{Miglio2012} have computed their mass estimates directly from the scaling relations. Going beyond such scaling relations by using stellar models would {\color{black}shift the mass estimates and potentially improve the agreement with the} values obtained from eclipsing binaries by \cite{Brogaard2012}. Thus, the asteroseismic analysis by \cite{2019ApJ...874..180M} {\color{black}based on stellar models} yields a mean mass estimate for the red giant branch stars in the cluster of $1.15\pm0.01\,\mathrm{M}_\odot$, which closely matches that by \cite{Brogaard2012}. This latter analysis notably also employs the two-term correction by BG14 (case b) but includes higher-{\color{black}degree} modes ($\ell=2$).

In short, the statistical errors of the reference age in Fig.~\ref{fig:SumAM_NGC6791} might be underestimated, while the reference mass might be subject to systematic errors.
Taking these circumstances into account, we note that the majority of the investigated surface correction relations lead to mass and age estimates that agree reasonably well with the literature. The case without any surface correction is clearly an example of an exception to this statement.

\begin{table}
    \centering
    \caption{Summary of the inferred mean mass and age ($\tau$) of the RGB stars in NGC~6791 based on the different surface correction relations that are investigated in this paper. The likelihood includes $\nu_\mathrm{max}$.}
    \label{tab:NGC6791}
    \begin{tabular}{lccccccccccccccccccccc} 
        \hline
        Surf. corr. & $\langle M \rangle \pm \sigma_{\bar{M}}$ & $\langle \tau \rangle \pm \sigma_{\bar{\tau}}$ [Myr] \\
        \hline
        K08 (a) & $1.161\pm0.011$ & $7157\pm 254$ \\ [2pt]
        K08 (b) & $1.170\pm0.010$ & $6893\pm219$ \\ [2pt]
        K08 (c) & $1.082\pm0.008$ & $9530\pm300$ \\ [2pt]
        S15 (a) & $1.094\pm0.010$ & $8930\pm340$ \\ [2pt]
        S15 (b) & $1.101\pm0.009$ & $8834\pm305$ \\ [2pt]
        S15 (c) & $1.110\pm0.012$ & $8141\pm411$ \\ [2pt]
        BG14 (a) & $1.236\pm0.014$ & $5466\pm227$ \\ [2pt]
        BG14 (b) & $1.126\pm0.010$ & $8122\pm300$ \\ [2pt] 
        NoSC (None) & $1.292\pm0.023$ & $4000\pm252$ \\ [2pt]
        \hline
\end{tabular}
\end{table}

By comparing Fig.~\ref{fig:SumAM_NGC6791} to Fig.~\ref{fig:SumAM} in Section~\ref{sec:NGC6819}, we note that the relative performance of the different surface correction relations is the same for both clusters. In other words, the ordering from the lowest to the highest mean mass or age is the same for both clusters. For instance, NoSC yields much higher masses and much lower ages and the remaining eight treatments of the surface effect do. {The same conclusion was drawn from the analysis of the eclipsing binaries in Section~\ref{sec:EB}.} This underlines that the use of improper surface correction relations does, indeed, lead to systematic errors in the obtained global parameters.

\begin{figure}
\centering
\includegraphics[width=1.00\linewidth]{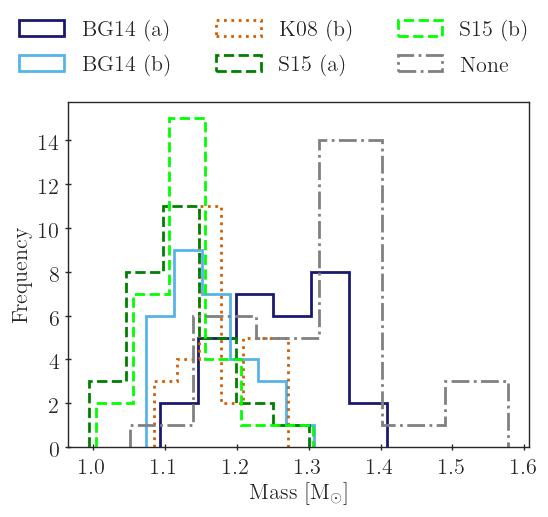}
\caption{Mass distributions for six of the nine treatments of the surface effect discussed in this paper {\color{black}for NGC~6791}. The remaining three approaches were excluded to avoid that the plot became too crowded. For each histogram, we employ six bins. These bins are therefore broader in mass for the cases with larger mass scatter. The likelihood includes $\nu_\mathrm{max}$.}
\label{fig:Mscat}
\end{figure}

{As can furthermore be seen from the error bars in Fig.~\ref{fig:SumAM_NGC6791}, the one-term correction by BG14 (case a) and the case without any surface correction relation yield broader mass distributions than the remaining seven treatments of the surface effect do --- the same is found for NGC~6819, although to a lesser extent. Since we are dealing with a cluster, this larger mass scatter indicates that the one-term correction by BG14 (case a) and the case without any surface correction relation are less reliable than the other approaches. We further illustrate this in Fig.~\ref{fig:Mscat}}

\begin{figure}
\centering
\includegraphics[width=1.00\linewidth]{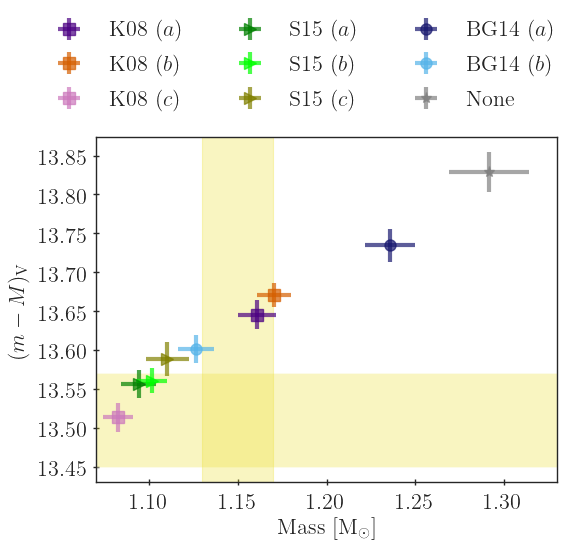}
\caption{As Fig.~\ref{fig:distance} but for NGC~6791. Here, we compare with the mass interval and the constraints on the distance modulus provided by Brogaard et al. (2012).}
\label{fig:distance6791}
\end{figure}

As in the case of NGC~6819, we have computed the distance modulus of the cluster. For this purpose, we used V-band magnitudes by \cite{Basu2011} in combination with the bolometric corrections by \cite{Casagrande2014}. With the exceptions of the one-term correction by BG14 (case a), cases a and b of K08, and NoSC, all treatments of the surface effect agree within $2\,\sigma$ with the value obtained from the study of eclipsing binaries by \cite{Brogaard2012} ($13.51\pm0.06$). We quantify this result in Fig.~\ref{fig:distance6791}. Once again, we note that the one-term correction by BG14 (case a) and NoSC perform worse than the remaining seven procedures. This notion is consistent with the results obtained in both Sections~\ref{sec:EB} and \ref{sec:NGC6819}. Only the two-term correction by BG14 recovers the stellar mass within $1\,\sigma$ \textit{in addition to} yielding a mean distance modulus that is consistent with \cite{Brogaard2012} within $2\,\sigma$.

\begin{figure}
\centering
\includegraphics[width=1.00\linewidth]{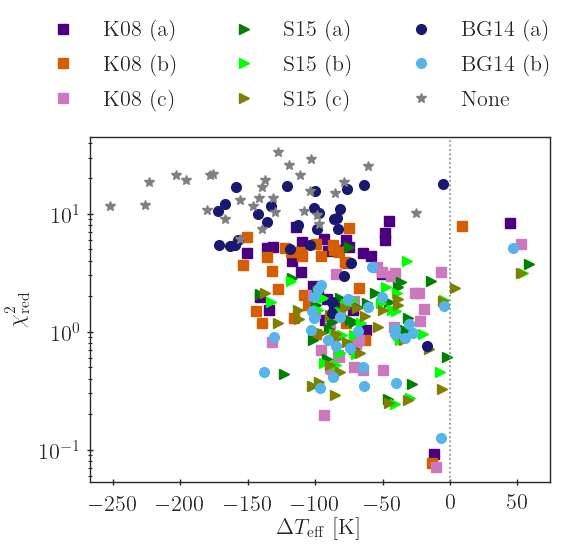}
\caption{The reduced $\chi^2$ for all stars in NGC~6791 as a function of the deviation between the measured and predicted effective temperature. Positive values for the temperature difference indicate that the {\color{black}spectroscopic constraint on the effective temperature} is higher than the {\color{black}inferred value}. The likelihood includes $\nu_\mathrm{max}$.}
\label{fig:test4_FID6701}
\end{figure}

While the inferred effective temperatures {\color{black}from \textsc{aims}} closely match {\color{black}spectroscopic constraints} in the case of NGC~6819, the inferred effective temperatures are too high for most cluster members in NGC6791, irrespectively of how we treat the surface effect.
Again, we find that the one-term correction by BG14 and NoSC lead to the worst goodness-of-fit scores, indicating that these approaches do not recover observations well. Meanwhile, in several cases, the other surface corrections lead to slight overfitting. We illustrate and quantify the discussed features in Fig.~\ref{fig:test4_FID6701}. 

On a star {\color{black}by} star basis, we find the mass estimates to be correlated with the attributed distance modulus, as mentioned in Section~\ref{sec:NGC6819}. We furthermore find the mass to be correlated with the offset between the effective temperature of the model and the spectroscopic constraint on $T_\mathrm{eff}$. We thus find stars that are assigned a lower mass to yield a higher offset in $T_\mathrm{eff}$. There is, meanwhile, no clear correlation between the offset in $T_\mathrm{eff}$ and the goodness of fit, if we solely consider models that share the same treatment of the surface correction relation. Overall, however, such a correlation exists, since the surface correction relations that yield the highest masses also lead to higher $\chi^2$- and BIC-values --- that is, case a of BG14 and NoSC. 

In this connection, we note that the evaluated $T_\mathrm{eff}$ is sensitive to the constraint on $\mathrm{[Fe/H]}$. The inferred values of $T_\mathrm{eff}$ and thereby of the distance moduli can thus be improved by changing the imposed metallicity of the cluster: changing the constraint on $\mathrm{[Fe/H]}$ by $0.1\,$dex increases the inferred temperature by roughly $50\,$K. However, we note that the mass and age estimates are likewise sensitive to this observational constraint. We can thus increase the mean mass and decrease the mean age estimates by assuming a systematically lower metallicity for the cluster members. An increase in the imposed metallicity by $0.1\,$dex thus decreases the mean mass estimate by roughly $0.02\,\mathrm{M_\odot}$ for all surface correction relations. In other words, in the presented scenario, the improvement of the distance modulus {\color{black}that is achieved by increasing the metallicity} comes at the cost of lowering the mean mass, {\color{black}which for most surface correction relations increases the tension with the dynamical constraints by \cite{Brogaard2012}.}

{\color{black}In the same manner, as regards the distance modulus, we can improve the agreement with \cite{Brogaard2012} by including alpha enrichment --- for all models presented in this paper, $\mathrm{[\alpha/Fe]}=0.0$ (cf. Section~\ref{sec:grid}). We thus repeated the analysis, adopting $\mathrm{[\alpha/Fe]}=0.1$. In this case, we find that all treatments of the surface effect but the one-term correction by BG14 (case a) and NoSC recover the mean distance modulus by \cite{Brogaard2012} within $2\,\sigma$. However, once again, this comes at the cost of lower mean masses and higher mean ages. The mean mass thus decreases by roughly $0.01-0.03\,\mathrm{M}_\odot$.
}

The effective temperature can also be shifted by choosing a different model atmosphere. To investigate this, we have repeated the analysis, substituting the semi-empirical model atmospheres by \cite{Vernazza1981} with those by \cite{KrishnaSwamy1966}. Doing so, we find that the offset in $T_\mathrm{eff}$ increases by roughly $50\,$K, leading to even higher model temperatures. As a result, none of the investigated treatments of the surface effect yields mean distance moduli that lie within $2\,\sigma$ of the constraints by \cite{Brogaard2012} when employing the model atmospheres by \cite{KrishnaSwamy1966}. Moreover, when using the model atmospheres by \cite{KrishnaSwamy1966}, only the cases a and b of S15 and case c of K08 lie within $1\,\sigma$ of the mass estimate by \cite{Brogaard2012}.

Finally, in connection with the discussion of Fig.~\ref{fig:distance6791}, we computed the mean distance moduli for all nine treatments of the surface effect based on the spectroscopic values for $T_\mathrm{eff}$ in combination with the stellar radii inferred by modelling. More specifically, we used the stellar radii from the models that enter Fig.~\ref{fig:distance6791}. We did so to further substantiate that the offset in the distance modulus is related to the mismatch in the effective temperature. Indeed, in this scenario, all nine treatments, except for case c of K08 and NoSC, lie within $1\,\sigma$ of the constraints by \cite{Brogaard2012}."


\section{Discussion} \label{sec:discussion7}

\begin{figure}
\centering
\includegraphics[width=1.00\linewidth]{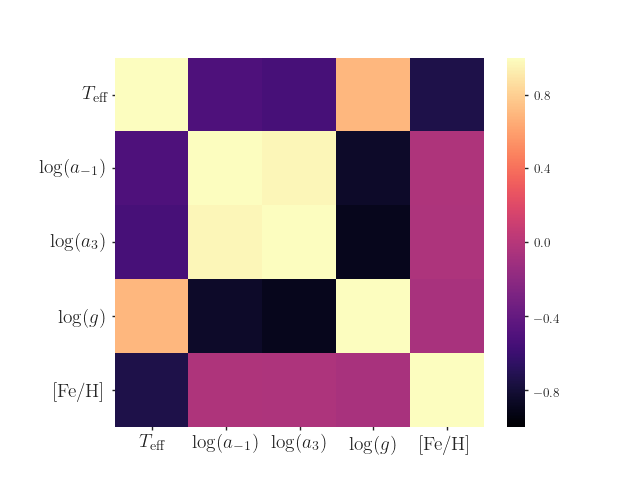}
\caption{Correlations between the surface correction relation parameters of BG14 (case b) and the global stellar parameters of the stars in NGC~6819 and NGC~6791 in combination with the eight eclipsing binaries in Section~\ref{sec:EB}. The correlation matrix has been computed based on the best-fitting models of each star. The analysis thus includes 57 red giant branch stars. The likelihood includes $\nu_\mathrm{max}$.}
\label{fig:MaCoall}
\end{figure}

Throughout the presented analysis of the eclipsing binaries, NGC~6819, and NGC~6791, the two-term correction by BG14 performs well. As argued by \cite{Ball2014} based on \cite{1990LNP...367..283G}, the underlying functional form is physically motivated, which may explain the success of this approach. Encouragingly, we furthermore find that the best-fitting values of both parameters involved in BG14 ($a_{-1}$ and $a_{3}$) are strongly correlated with the global stellar parameters, predominantly $\log g$, of the best-fitting models. Thus, the inferred surface effect evolves in a predictable manner through the HR diagram. This is illustrated in Fig.~\ref{fig:MaCoall}.

\begin{figure}
\centering
\includegraphics[width=1.00\linewidth]{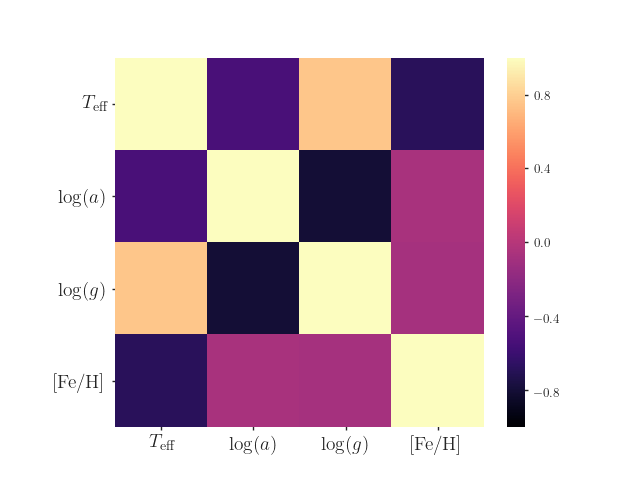}
\includegraphics[width=1.00\linewidth]{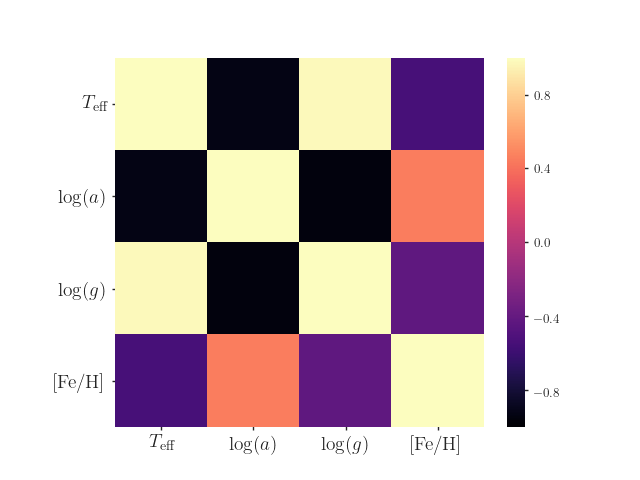}
\caption{As Fig.~\ref{fig:MaCoall} but for K08 (b). The upper panel is based on the global stellar parameters of the stars in NGC~6819 and NGC~6791 in combination with the eight eclipsing binaries in Section~\ref{sec:EB}. For the lower panel, we have only included the stars from NGC~6819.}
\label{fig:MaCoall2}
\end{figure}

As regards K08 and S15, we likewise find the involved parameters ($a$, $b$, $\alpha$, and $\beta$) to be correlated with the global stellar parameters, when {\color{black}$a$, $b$, $\alpha$, and $\beta$} are kept free. 

We note that the discussed correlations partly reflect the intrinsic correlations between the global stellar parameters, say, between $T_\mathrm{eff}$ and $\log g$. These intrinsic correlations come about as red giant branch stars in any given cluster will share similar properties --- indeed, this notion underlies the analysis presented in the previous sections. In other words, the intrinsic correlations reflect a selection bias. Notably, repeating the analysis solely based on a single cluster, the results occasionally point towards correlations between the free surface correction relation parameters and $\mathrm{[Fe/H]}$. If this result were to be genuine, it would be quite intriguing, since such correlations are not taken into account by Eqs~(\ref{eq:loga}), (\ref{eq:logb}), (\ref{eq:logalpha}), and (\ref{eq:logbeta}). However, these correlations are washed away when including more data. This is exemplified in Fig.~\ref{fig:MaCoall2}: based on NGC~6819 alone, $a$ seems to be strongly correlated with $\log g$, $T_\mathrm{eff}$, and $\mathrm{[Fe/H]}$ when considering case b of K08 (lower panel). The picture changes when including an additional cluster and the eclipsing binaries (upper panel). Weaker correlations with $T_\mathrm{eff}$ and $\mathrm{[Fe/H]}$ are obtained. {This is not to say that the metallicity might not play a role for the parameters involved in the surface correlation relations. However, while correlations between the surface correction relation parameters and the global stellar parameters might express dependencies on these global parameters, these correlations also indirectly express a selection bias.}

Moreover, our results suggest that Eqs (\ref{eq:loga}), (\ref{eq:logalpha}), and (\ref{eq:logbeta}) systematically underestimate the best-fitting values of $a$, $\alpha$, and $\beta$, while Eq.~(\ref{eq:logb}) overestimates $b$. This was to be anticipated, as \cite{Sonoi2015} have calibrated these relations mainly based on main-sequence stars, for which $T_\mathrm{eff}$ lies above $6000\,$K. The population that is studied in the present paper thus strongly differs from that, based on which \cite{Sonoi2015} derived their expressions for $a$, $b$, $\alpha$, and $\beta$.

In order to further quantify these statements, one may attempt to calibrate the parameters in K08 and S15 based on a large sample of well-constrained stars, such as eclipsing binaries. {The use of such benchmark targets allows adjusting the parameters based on well-established non-seismic constraints (cf. Fig.~\ref{fig:MRbinaR}).}
However, since the obtained values may, to a large extent, reflect other input physics, such as the chosen $T(\tau)$ relation, and since there is no physical justification for the functional form of K08 and S15, a calibration of these surface correction relations is not guaranteed to be widely applicable. Indeed, as discussed by \cite{Joergensen2019}, any attempt to establish a global calibration of K08 and S15 may be subject to a selection bias. {In other words, the parameters of surface correction relations encode physical inadequacies of the stellar models, and they thus reflect the associated input physics as well as the global stellar parameters, since the mentioned inadequacies are sensitive to these properties. It thus stands to reason that a proper calibration of K08 and S15 must be performed based on a suitable set of benchmark stars, every time the input physics is significantly altered, and every time a yet uncovered region of the HR diagram is explored.}

\begin{figure}
\centering
\includegraphics[width=1.00\linewidth]{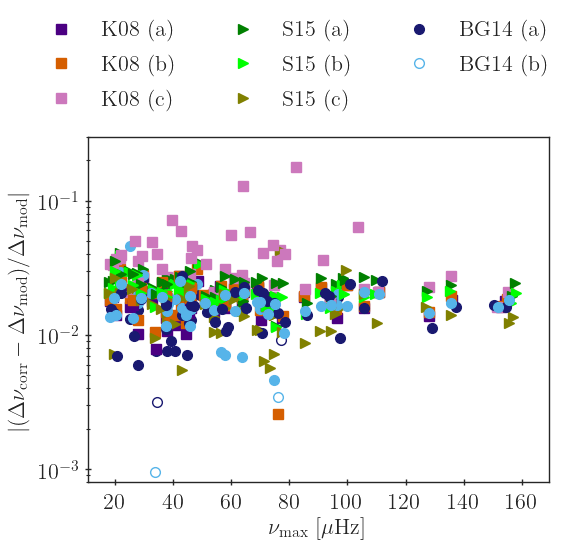}
\caption{Relative difference in the large frequency separation between that obtained {\color{black}from} corrected frequencies ($\Delta \nu_\mathrm{corr}$) and that obtained from uncorrected frequencies ($\Delta \nu_\mathrm{mod}$). The plot is based on all 57 red giant branch stars presented in this paper. One outlier with a relative difference below $10^{-3}$ were removed from the plot for clarity --- it used case a of K08. We note that $\Delta \nu_\mathrm{corr}<\Delta \nu_\mathrm{mod}$, in nearly all cases. {\color{black}Exceptions only occur for BG14 and have been indicated with open symbols}. The averages constitutes unweighted means taken over the model frequencies that correspond to the observed modes. }
\label{fig:DNUtest}
\end{figure}

{While we do not dive further into the coefficients and exponents that enter the surface correction relations, it is worth taking a closer look at the consequences of using the surface correction relations for the predicted physical stellar properties. We thus note that the relative shift in the large frequency separation is constant across all values of $\nu_\mathrm{max}$. This is illustrated in Fig.~\ref{fig:DNUtest} \citep[see also][]{Rodrigues2017}. For nearly all cases, the large separation is lower when including a correction of the surface effect\footnote{The very few exceptions all employ the one- or two-term correction by BG14. Note that we indirectly enforced this behaviour for K08 and S15 by imposing priors on the involved parameters (cf. Table~\ref{tab:param}).}.

Case c of K08 constitutes a notable exception to this rule and deviates by one order of magnitude in the size of the relative shift in $\Delta \nu$ from the other surface correction relations. The fact that this correction significantly differs in the predicted surface effect is consistent with the notion that case c of K08 leads to lower mass estimates than the remaining surface correction relations do, including the cases a and b of K08, as shown in Figs~\ref{fig:SumAM} and \ref{fig:SumAM_NGC6791}.}

\section{Conclusion} \label{sec:conclusion}

In this paper, we perform asteroseismic analyses of several red giant branch stars imposing different surface correction relations \cite[cf.][]{Kjeldsen2008,Ball2014,Sonoi2015}. We demonstrate that the use of each of these surface correction relations biases the inferred stellar properties. In accordance with the studies of main-sequence stars by \cite{Ball2017} and \cite{Nsamba2018}, our results thus show that the use of different surface correction relations leads to systematic offsets in the derived global stellar properties, including stellar ages, masses, and distances.

In Section~\ref{sec:EB}, we present an asteroseismic analysis of eight giants that are situated in eclipsing binaries. From comparisons with the results of dynamical studies, we conclude that we are able to recover the global stellar parameters equally well independently of the employed surface correction relation if we impose sufficiently informative non-seismic constraints. Even with such non-seismic constraints, however, there is a scatter between the inferred stellar properties that are favoured by the different surface correction relations. Throughout most of this paper, we ensure that the constraints on the individual frequencies dominate the likelihood in order to discriminate between the different surface correction relations. Meanwhile, we also demonstrate how the use of different additional constraints that complement the individual frequencies reduces the impact of the chosen surface correction relation. When including dynamical constraints on the stellar radius or constraints on $\nu_\mathrm{max}$, most treatments of the surface effect match the dynamical constraints on the masses of the binary members within $2\,\sigma$ and $3\,\sigma$, respectively. 
Without any of these constraints, however, only the two-term correction by \cite{Ball2014} is able to recover the dynamical measurements of the stellar masses and radii for all stars in the sample within $3\,\sigma$. Especially, the approach of ignoring the surface effect altogether yields poor results. Moreover, in several cases, the power-law description by \cite{Kjeldsen2008} leads to un-physically large frequency offsets in the un-corrected frequencies. We attribute this behaviour to the fact that there is no physical justification for the power-law correction beyond the calibration of the involved coefficient and exponent.

Based on our analysis of eclipsing binaries, we thus generally recommend to include available non-seismic constraints in asteroseismic analyses, to lower any systematic errors that might be introduced by the chosen treatment of the surface effect. Surveys and missions, such as Gaia, provide such constraints.

In Sections~\ref{sec:NGC6819} and \ref{sec:NGC6791}, we address a large sample of red giant branch stars from two open clusters and investigate the attributed mean properties of the populations. Through this analysis, we illustrate that the use of different surface correction relations biases mass, age, and distance estimates for the cluster members.
In other words, improper treatments of the surface effect lead to systematic errors in the obtained global parameters. In this connection, we note that the different surface correction relations perform in the same manner relative to each other for both clusters. For instance, for both clusters, the one-term correction relation by \cite{Ball2014} yields higher mean mass and lower mean age estimates than the remaining surface correction relations do. Furthermore, when using the surface correction relation from \cite{Sonoi2015}, one on average obtains lower mass estimates than when employing the two-term correction by \cite{Ball2014}. We also note that the obtained results are inconsistent with existing constraints on the mass and age of the clusters when no surface correction is imposed. These conclusions are consistent with our analysis of the eclipsing binaries in Section~\ref{sec:EB}.

Finally, we illustrate that the attempt to calibrate a surface correction relation based on a sample of target stars introduces correlations in the derived coefficients and exponents, reflecting the underlying sample. We note that this agrees with the results obtained by \cite{Joergensen2019}, who show that the underlying sample leads to a selection bias. In combination with the results presented by \cite{Joergensen2019}, our results thus discourage from the use of calibrated surface correction relations in connection with targets that do not closely resemble the stars, based on which the relations were calibrated. {Overall, our results are thus not in favour of using the surface correction relations by \cite{Kjeldsen2008} and \cite{Sonoi2015} without a proper calibration of the parameters involved. Such a calibration must be based on the input physics of the employed models as well as the global parameters of the target star.

The surface correction relation by \cite{Ball2014}, on the other hand, does not rely on any such calibration. Nevertheless, our results discourage from the use of the one-term correction by \cite{Ball2014}, since the use of this surface correction relation does not recover the correct stellar properties and leads to models with a poor goodness-of-fit score. Moreover, in the presented analysis of the two clusters, the one-term correction by \cite{Ball2014} leads to the largest mass scatter among the investigated surface correction relations. This finding indicates that the one-term correction by \cite{Ball2014} is less reliable than the other surface correction relations. On the other hand, the two-term correction by \cite{Ball2014} performs very well throughout the analysis presented in this paper. It recovers the classical constraints on the binaries and both clusters, and it leads to models with an excellent goodness-of-fit score.

Neglecting the surface effect altogether is demonstrably not a viable strategy, as it significantly skews the obtained estimates for the global stellar parameters. When addressing the binary members in Section~\ref{sec:EB}, the approach of neglecting the surface effect altogether thus yields parameter estimates that strongly deviate from the dynamical constraints, whether or not we include additional (non-seismic) constraint to complement the individual frequencies. This approach also leads to larger mass scatter on a star {\color{black}by} star basis when addressing the two clusters.}

The surface effect refers to a frequency offset that partly arises from an incomplete description of the outermost superadiabatic layers of stars with convective envelopes. When addressing the surface effect, it is, therefore, worth noting that the underlying structural shortcomings affect the outer boundary conditions for the evaluated interior equilibrium structure \citep[e.g.][]{Kippenhahn}. As a result, the surface effect is a symptom of a model inadequacy that also affects the predicted stellar evolution tracks. This is demonstrated by e.g. \cite{Mosumgaard2020} based on three-dimensional magneto-hydrodynamic simulations by \cite{Magic2013} and based on the method by \cite{Joergensen2018}. Surface correction relations do not account for such changes in the global stellar parameters. In other words, surface correction relations do not deal with all aspects that are associated with the structural shortcomings that give rise to the surface effect. However, to address this issue, one needs to go beyond some of the assumptions that enter state-of-the-art stellar models as shown by e.g. \cite{Joergensen2018} and \citep{Mosumgaard2018}. This is beyond the scope of the present paper. In this paper, we rather demonstrate that the correct treatment of the surface effect is crucial for the outcome of asteroseismic analyses when employing state-of-the-art stellar evolution codes.


\section*{Acknowledgements}

We thank {\color{black}Karsten Brogaard,} J. Ted Mackereth, and Martin B. Nielsen for their input and useful discussions.
{\color{black}We furthermore acknowledge the diligent and helpful feedback of our anonymous referee.}
The research leading to this paper has received funding from the European Research Council (ERC grant agreement no.772293 for the project ASTEROCHRONOMETRY).
G.B. acknowledges funding from the SNF AMBIZIONE grant No 185805 (Seismic inversions and modelling of transport processes in stars) and funding from the European Research Council (ERC) under the European Union's Horizon 2020 research and innovation programme (grant agreement No 833925, project STAREX). {\color{black}R.A.G acknowledges the support received from the PLATO CNES grant.}




\bibliographystyle{mnras}
\bibliography{manual_refs,mendeley_export}








\bsp    
\label{lastpage}
\end{document}